\documentclass[journal]{IEEEtran}
\ifCLASSINFOpdf
\else
   \usepackage[dvips]{graphicx}
\fi
\usepackage{url}

\usepackage{algorithmic,algorithm,xpatch}
 \usepackage{setspace}

\xpatchcmd{\algorithmic}{\setcounter}{\algorithmicfont\setcounter}{}{}
\providecommand{\algorithmicfont}{}
\providecommand{\setalgorithmicfont}[1]{\renewcommand{\algorithmicfont}{#1}}

\usepackage{algorithm}
\usepackage{algorithmic}
\usepackage {hyperref}
\usepackage{amsthm}
\usepackage{siunitx}
\usepackage{subfig}
\usepackage{mathtools}
\usepackage{cite}

\usepackage{caption}

\theoremstyle{remark}
\newtheorem*{remark}{Remark}

\newtheorem*{notation}{\textbf{Notation}}

\usepackage{graphicx}
\usepackage{cite}
\usepackage{tikz}
\usetikzlibrary{arrows}
\usepackage{verbatim}
\usepackage{amsmath,amssymb,mathtools,bm,etoolbox}
\usepackage{scalerel}

\newcommand\sbullet[1][.5]{\mathbin{\vcenter{\hbox{\scalebox{#1}{$\bullet$}}}}}

\makeatletter
\patchcmd{\@maketitle}
  {\addvspace{0.5\baselineskip}\egroup}
  {\addvspace{-1\baselineskip}\egroup}
  {}
  {}
\makeatother
\DeclarePairedDelimiterX\set[1]\lbrace\rbrace{\def\given{\;\delimsize\vert\;}#1}
\usepackage{amssymb}

\newcommand*{\field}[1]{\mathbb{#1}}%

\begin{document}

\raggedbottom

\title{\LARGE{Underdetermined Blind Identification via $k$-Sparse Component Analysis: RANSAC-driven Orthogonal Subspace Search}}

\author{Ehsan Eqlimi*, \IEEEmembership{Member, IEEE}, Bahador Makkiabadi*, \IEEEmembership{Member, IEEE}, Mayadeh Kouti, Ardeshir Fotouhi, Saeid Sanei, \IEEEmembership{Senior Member, IEEE}
\thanks{

 Ehsan Eqlimi is with WAVES Research Group, Department of Information Technology, Ghent University, Ghent, Belgium, and also Department of  Quality  and Process, OLV hospital, Aalst, Belgium

 Ehsan Eqlimi, Bahador Makkiabadi, and Ardeshir Fotouhi are with  Department of Medical Physics and
Biomedical Engineering, School of Medicine, Tehran University of Medical Sciences, Tehran, Iran

Mayadeh Kouti is with Department of Electrical Engineering, Shahid Chamran University of Ahvaz, Ahvaz, Iran

Saeid Sanei is with School of Science and Technology, Nottingham Trent
University, Nottingham, UK

 *\href{mailto:ehsan.eqlimi@outlook.com}{ehsan.eqlimi@outlook.com} $\&$ *\href{mailto:b-makkiabadi@sina.tums.ac.ir }{b-makkiabadi@sina.tums.ac.ir} (corresponsing authors)

The Matlab\textregistered~codes implementing the algorithm are
publicly accessible on GitHub (\url{https://github.com/EhsanEqlimi/k-SCA-UBI-Eusipco2023}).  

}}

%

\maketitle
\thispagestyle{empty} 

\begin{abstract}


Two primary families of methods exist for underdetermined blind identification (UBI) based on the sparsity of the source matrix: sparse component analysis (SCA) and $k$-SCA. SCA assumes one active source at each time instant, while $k$-SCA allows for varying numbers of active sources represented by $k$. However, existing $k$-SCA methods, which claim to solve UBI problems by accommodating $k$-sparse sources, predominantly rely on $1$-sparse sources, limiting their effectiveness in real-world scenarios with high noise levels.

In this paper, we propose an effective and computationally less complex approach for UBI, specifically focusing on the challenging case when the number of active sources is equal to the number of sensors minus one ($k=m-1$). Our approach overcomes limitations by using a two-step scenario: (1) estimating the orthogonal complement subspaces of the overall space and (2) identifying the mixing vectors. We present an integrated algorithm based on the Gram-Schmidt process and random sample consensus (RANSAC) method to solve both steps. Experimental results using simulated data demonstrate the superior effectiveness of our proposed method compared to existing algorithms.
\end{abstract}


\IEEEpeerreviewmaketitle

\section{Introduction}
\label{IntroSec}
\IEEEPARstart{U}{nderdetermined} blind source separation (UBSS) aims at separating the source signals from their instantaneous linear mixtures when there are more sources than sensors. UBSS is a challenging problem since its mixing matrix is not invertible \cite{bofill2001underdetermined}. A two-step approach is often adopted to solve UBSS: (1) underdetermined blind identification (UBI) to identify the mixing matrix and (2) recovering the source matrix. In this paper, precise identification of the mixing matrix is considered, which is more challenging than source recovery. Sparse component analysis (SCA) is widely used to address the UBSS problem if the sources are sparse \cite{bofill2001underdetermined,georgiev2005sparse,li2006underdetermined}. A key issue in SCA is the identifiability condition leading to the identification of a unique mixing system (up to scaling and permutation ambiguities). There are two scenarios for SCA identifiability. First, the probabilistic approach mostly relies on a prior distribution for the values and locations of non-zero elements of the source matrix \cite{gribonval2015sparse}. Second, the deterministic approach tries to impose definite conditions on the source and mixing matrices. This paper relies on the latter approach \cite{georgiev2005sparse,aharon2006uniqueness}.

One of the key conditions for deterministic SCA identifiability is the number of non-zero elements in each column of the source matrix, referred to as the sparsity level, $k$. It has been shown that the lowest sparsity case corresponds to $k = m-1$, where $m$ represents the number of sensors, enabling matrix identification under the $k$-SCA identifiability conditions~\cite{georgiev2005sparse}. Despite the explicit proof provided, the proposed UBI algorithm is characterized by its generality and implicit nature. While several UBI algorithms have been proposed based on $k$-SCA conditions~\cite{washizawa2006line,he2009k,wen2014new,eqlimi2015multiple,he2016underdetermined}, they do not perform well when the columns of the source matrix share the same sparsity levels. Additionally, most of these algorithms are not robust to noise that replaces inactive sources. However, we have proposed a few approaches to improve the upper bound for the maximum possible number of non-zero sources in previous works~\cite{eqlimi2015efficient,eqlimi2015multiple,eqlimi2019novel}. Nonetheless, these algorithms are not robust to noise and are dependent on multiple threshold definitions.

Motivated by the $k$-SCA theorem~\cite{georgiev2005sparse} and the aforementioned limitations, this paper presents an integrated, efficient, and straightforward algorithm specifically designed for the small-scale UBI scenario when $k = m-1$. To estimate the representative subspace of $k$-dimensional subspaces, which corresponds to the complementary orthogonal subspace, we propose an algorithm that combines the Gram-Schmidt process and the random sample consensus (RANSAC) method~\cite{fischler1981random}. This algorithm effectively addresses both steps of the problem, namely, estimating the underlying subspaces and identifying the mixing matrix. We conducted several simulation experiments to demonstrate the effectiveness of our proposed algorithm.

\section{Problem Statement}
\label{ProbStatSec}
The linear instantaneous mixing system of UBSS, in the noiseless case, can be expressed as:

\begin{equation}
\label{UBSSEQ}
\textbf{X}=\textbf{A}\textbf{S},\quad \textbf{A} \in \mathbb{R}^{m \times n}, \quad \textbf{S} \in\mathbb{R}^{n \times T},
\end{equation}

\noindent where $T$ represents the number of time (or transformed domain) samples, and the $m$ mixture signals (rows of the mixture matrix, $\textbf{X}$) result from the linear and instantaneous mixing of $n > m$ unknown source signals (rows of the source matrix, $\textbf{S}$) with an unknown mixing matrix, $\textbf{A}$. For our purposes, we present Equation (\ref{UBSSEQ}) in vector form as follows:

\begin{equation}
\label{VectorviseUBSSEq}
 \textbf{x}(t)= \sum_{q=1}^{n}{\textbf{a}_q s_q(t)} \quad t= 1,\dots,T,
\end{equation}

\noindent where $\textbf{x}(t) \in \mathbb{R}^{m \times 1}$ represents the $m$ mixture values at time instant $t$, $s_q(t) \in \mathbb{R}^{1 \times 1}$ denotes the $q^{th}$ source value at time instant $t$, and $\textbf{a}_q \in  \mathbb{R}^{m \times 1}$ represents the $q^{th}$ column of the mixing matrix. By increasing the number of zeros in $\textbf{S}$ at each time instant, the number of corresponding columns in $\textbf{A}$ to $\textbf{x}(t)$ can be reduced. The complexity of identifying $\textbf{A}$ varies depending on the sparsity level. The mixture signals associated with the same contributing columns of $\textbf{A}$ can be considered within a $k$-dimensional subspace. In general, the total number of joint subspaces is equal to $c={ n \choose k}$. According to the Gram-Schmidt theorem, if $\{x_1,...,x_k\}$ is a linearly independent list of vectors, there exists an orthonormal list $\{e_1,...,e_k\}$ in the space such that $\textrm{span}(x_1, x_2,..., x_k)= \textrm{span}(e_1, e_2,..., e_k)$. To identify the mixing columns or the basis columns of the subspaces, we can employ the concept of orthogonal complement subspaces. In the next section, we will explain the proposed algorithm for identifying these subspaces and using them to determine the mixing matrix.
\begin{notation}
\small{
The operators $[\sbullet,\sbullet]$, $\langle \sbullet ,\sbullet \rangle$, and $\left| \sbullet \right|$ represent the horizontal concatenation of two matrices, the inner product of two vectors, and the $\ell_2$-norm, respectively. A vector $\textbf{s}$ is considered $k$-sparse if it has $k$ non-zero entries, denoted as $\left| \textbf{s}\right|_0=k$. Let $\mathbb{J}={j_1,j_2,\dots,j_b}$ be the indices of the $b$ selected columns of $\textbf{X}$. The submatrix of $\textbf{X}$ formed by selecting all rows and columns from ${j_1,j_2,\dots,j_b}$ is denoted as $\textbf{X}(:,\mathbb{J})$. A similar notation is used to indicate the selection of a matrix from a tensor, denoted as $\mathcal{X}(:,:,\mathbb{J})$. 

}
\end{notation}

\section{Proposed Algorithm}
\label{PropAlgSec}
In this paper, we propose a RANSAC-based subspace search algorithm to address the UBI problem when $k \leq m-1$. The framework overview is depicted in Fig.~\ref{OverviewFig}. The algorithm comprises two stages, each represented by a separate block in Fig.~\ref{OverviewFig}. The objective of the first stage is to identify the orthogonal complement subspaces (OCSs) through a subspace search algorithm that combines the Gram-Schmidt process with the RANSAC method. This stage aims to estimate the underlying subspaces that contribute to the observed mixture signals. In the second stage, a similar procedure is employed to identify the mixing matrix. The goal here is to determine the columns of the mixing matrix that correspond to the identified OCSs obtained from the first stage. Overall, our algorithm offers a comprehensive approach for solving the UBI problem, leveraging the power of RANSAC and the Gram-Schmidt process in both stages of the algorithm.


\tikzstyle{int}=[draw, fill=blue!20, minimum size=2em]
\tikzstyle{init} = [pin edge={to-,thin,black,}]

\begin{figure}[H]

\begin{tikzpicture}[node distance=3.7cm,auto,>=latex']

    \tikzstyle{every node}=[font=\footnotesize]

    \node [int,scale=1] (a) {\begin{tabular}{c} Cluster into
$c$ groups  \end{tabular} };
    \node (b) [left of=a,node distance=2.5cm, coordinate] {a};
    \node [int] (c) [right of=a] {\begin{tabular}{c} Cluster into
$n$ groups   \end{tabular}};
    \node [coordinate] (end) [right of=c, node distance=2.5cm]{};
    \path[->] (b) edge node {$\textbf{X}$} (a);
    \path[->] (a) edge node {$\mathcal{P}$} (c);
    \draw[->] (c) edge node {$\hat{\textbf{A}}$} (end) ;
  
\end{tikzpicture}
\captionsetup{font=small}

 \caption{The block diagram illustrates the proposed algorithm for identifying the mixing matrix, $\textbf{A} \in \mathbb{R}^{m \times n}$. In the first stage, the output consists of $c$ orthogonal complement subspaces (OCSs) of the underlying subspaces, denoted as $\mathcal{P} \in \mathbb{R}^{m \times b \times c}$, where $b=m-k$.}
\label{OverviewFig}

\end{figure}

\subsection{Identifying the orthogonal complement subspaces}
The proposed algorithm aims to identify the orthogonal complement subspaces of a $k$-dimensional space given the input $\textbf{X}$ and the value of $k$. In this problem, each column of $\textbf{X}$ is assumed to lie approximately on the union of multiple $k$-dimensional linear subspaces spanned by corresponding columns of the mixing matrix $\textbf{A}$. Therefore, the first step in solving this problem is to estimate these subspaces.

To accomplish this, a robust model is fitted to the columns of $\textbf{X}$ using the RANSAC algorithm. RANSAC is an iterative approach that allows for the extraction of model parameters from observed data, even in the presence of outliers~\cite{fischler1981random}. By iteratively fitting the model to the observed data, the algorithm can robustly estimate the subspaces that best represent the columns of $\textbf{X}$.


The RANSAC algorithm is employed to address the UBI problem by utilizing fitting, distance, and degenerate functions (equations (\ref{EqFit})-(\ref{EqDis})). Given that there are $c=$ ${n}\choose{k}$ subspaces, each spanned by $k$ basis vectors, it is necessary to repeat the RANSAC process at least $c$ times. This repetition is crucial to ensure the estimation of all $c$ subspaces and the successful clustering of an adequate number of data points in each iteration.



We utilize the Gram-Schmidt process to design the fitting function, which is used to define a model based on the observed data. The fitting function performs model fitting by selecting $l$ randomly sampled data points and applying the following procedure:

\begin{equation}
\label{EqFit}
\hat{\textbf{P}}=\textbf{I}_{m \times m}- \textrm{GS}([\textbf{x}_1,...,\textbf{x}_l])*\textrm{GS}([\textbf{x}_1,...,\textbf{x}_l])^\textrm{T},
\end{equation}
\noindent
where $\textrm{GS}$ represents the Gram-Schmidt process, $\textbf{I} \in \mathbb{R}^{m \times m}$ denotes the identity matrix, $\textbf{X}_{sel} = [\textbf{x}_1, ..., \textbf{x}_l] \in \mathbb{R}^{m \times l}$ represents a submatrix of $\textbf{X}$ containing $l$ randomly selected columns, and $\hat{\textbf{P}} \in \mathbb{R}^{m \times m}$ represents the pseudo orthogonal complement subspace of the selected data points. The Gram-Schmidt process is a method used to construct an orthogonal basis from a set of linearly independent vectors~\cite{bjorck1994numerics}.


In order to ensure that the randomly selected columns are not in a degenerate configuration, we use a degenerate function, defined as follows:

\begin{equation}
\label{EqDeg}
r=\textrm{rank}(\textbf{X}_{sel}),
\end{equation}
\noindent
where the degenerate function measures the rank of the selected columns in $\textbf{X}_{sel}$, denoted as $\textrm{rank}(\textbf{X}_{sel})$. If the rank is less than $l$, it indicates that $\textbf{X}_{sel}$ is in a degenerate configuration, as it does not contain $l$ linearly independent columns. It is important to note that in Algorithm \ref{Alg 1}, $l$ is equal to the sparsity level ($l=k$), whereas in Algorithm \ref{Alg 3} discussed in subsection \ref{IMM}, $l$ is set to $m-1$.

The distance function, also known as the score function, plays a crucial role in evaluating the quality of a candidate solution. It provides the indices of inlier data points and measures the distance between $\textbf{X}(:, \mathbb{J})$ and $\hat{\textbf{P}}$, where $\mathbb{J}\subseteq \set{t \in \field{N} \given 1\leq t \leq T }$ is a subset of indices corresponding to the current set of inliers (refer to Algorithm~\ref{Alg 1}). Essentially, the distance function quantifies how closely a given candidate aligns with the subspace $\hat{\textbf{P}}$. It accomplishes this by computing the projection of a data vector $\textbf{x}_j=\textbf{X}(:, j)$ (where $j \in \mathbb{J}$) onto $\hat{\textbf{P}}$. The projected vector, denoted as $\textrm{proj}_{\hat{\textbf{P}}}^{{\textbf{x}}_j}$, represents the closest vector in $\hat{\textbf{P}}$ to $\textbf{x}_j$. Thus, the minimum distance between $\textbf{x}_j$ and the subspace $\hat{\textbf{W}} = \hat{\textbf{P}}^{\perp}$ can be expressed as $||\textrm{proj}_{\hat{\textbf{P}}}^{\textbf{x}_j}||$. 

Considering that $\hat{\textbf{P}}$ is the orthogonal complement subspace spanned by $l$ basis vectors, the distance function employs the projection onto the row space to define the distance as follows:

\begin{equation}
\label{EqDis}
d_j=\sum_{i=1}^{m}{\langle \hat{\textbf{P}}(i,:)^\textrm{T},\textbf{x}_j \rangle^2},
\end{equation}
\noindent
where $\hat{\textbf{P}}(i,:)$ represents the $i^{th}$ row of $\hat{\textbf{P}}$. If the computed distance $d_j$ is smaller than the predefined thresholds ($Th_1$ or $Th_2$), it indicates that the selected data point $\textbf{x}_j$ is closely aligned with $\hat{\textbf{W}}$ and is considered an inlier. Smaller distance values correspond to higher score values, indicating a stronger alignment.

The algorithm block diagram is illustrated in Fig.~\ref{BDAlg1}. After obtaining the inlier data through RANSAC in each iteration, singular value decomposition (SVD) is applied to decompose the inlier data and estimate the orthogonal complement subspaces. The candidate set of inlier indices is then updated by removing the indices of the identified inlier data points from $\mathbb{J}$. The pseudo-code for the algorithm is provided in Algorithm~\ref{Alg 1}.

\begin{figure}[H]
\begin{tikzpicture}[node distance=3cm,auto,>=latex']

    \tikzstyle{every node}=[font=\scriptsize]

    \node [int]
     (a) { \begin{tabular}{c} RANSAC using \\
 Eq.~(\ref{EqFit})-(\ref{EqDis})
\end{tabular}};
    \node (b) [left of=a,node distance=1.7cm, coordinate] {a};
    \node [int] (c) [right of=a] {\begin{tabular}{c} Find OCSs via \\ SVD  of $\textbf{X}(:,\mathbb{Y})$ \end{tabular}};
    \node [coordinate] (end) [right of=c, node distance=2cm]{};
       \node [int]   (d) [right of=c] {\begin{tabular}{c} Update $\mathbb{J}$ by\\ $\mathbb{J} \leftarrow \mathbb{J} \backslash \mathbb{Y}$ \end{tabular}};
    \path[->] (b) edge node {\begin{tabular}{c}   $\textbf{X}$ \end{tabular} } (a);
    \path[->] (a) edge node {$\mathbb{Y}$} (c);
    \draw[->] (c) edge node {$\mathcal{P}$} (end) ;
            \draw [->] (d) -- ++ (0,-1) -| node [pos=0.75] {$\mathbb{J}$} (a);
		 
\end{tikzpicture}
 
\captionsetup{font=small}

\caption{The block diagram in Algorithm~\ref{Alg 1} illustrates the process of finding the orthogonal complement subspaces (OCSs), denoted as $\mathcal{P}$. The algorithm iteratively repeats this process until all the OCSs have been estimated.}
%

\label{BDAlg1}
\end{figure}

\setalgorithmicfont{\small}

\begin{algorithm}
\captionsetup{font=small}

\caption{Identification of the orthogonal complement subspaces based on
RANSAC-based subspace search} 
\label{Alg 1}
\begin{algorithmic}
   \STATE \textbf{Initialization:} $Th_1$ (distance threshold for RANSAC).
   \STATE \textbf{Input:} $\textbf{X}$ (mixture matrix), $n$ (number of sources), and $l=k$ (sparsity level).
      \STATE \textbf{Output:} $\mathcal{W} \in{\mathbb{R}^{m \times k \times c}}$ and $\mathcal{P} \in{\mathbb{R}^{m \times (m-k) \times c}}$.
 \STATE \textbf{1:} Assign an initial set of inlier indices, $\mathbb{J}\leftarrow \{1, \dots, T\}$ ($\set{j \in \field{N} \given 1\leq j \leq T }$).
 \FOR{$i=1$ to $c={{n}\choose{k}}$} \STATE { \textbf{2:} Apply RANSAC method to $\textbf{X}(:, \mathbb{J})$ following equations (\ref{EqFit})-(\ref{EqDis}) to obtain the indices set of the inlier points $\leftarrow \mathbb{Y} \subset \mathbb{J}$.}
 \STATE \textbf{3:} Decompose $\textbf{X}(:, \mathbb{Y})$ to $\textbf{U}\boldmath{\Sigma}\textbf{V}^\textrm{T}$ using SVD and find the subspace, $\mathcal{W}(:,:,i) \leftarrow  \textbf{U}(:,1:k)$ and orthogonal complement space,
$\mathcal{P}(:,:,i) \leftarrow  \textbf{U}(:,k + 1:end)$.
\STATE \textbf{4:} Update the set $\mathbb{J}$ by removing the elements belonging to $\mathbb{Y}$ from $\mathbb{J}$, i.e., $\mathbb{J} \leftarrow \mathbb{J} \backslash \mathbb{Y}$ ($\{j\in\mathbb{J}~|~j\not\in \mathbb{Y}\})$ where ${( \sbullet \backslash \sbullet })$ stands for set subtraction.

  \ENDFOR
\end{algorithmic}
\end{algorithm}

\begin{remark}
\small{
The number of iterations, $N$, plays a crucial role in achieving the desired outcome in the RANSAC process. In each trial, there is a probability $\omega$ of selecting an inlier. The probability of selecting a sample subset with outliers in all $N$ trials can be calculated as $(1 - \omega^l)^N$, where $l$ is the minimum number of samples required for fitting. The success probability of a RANSAC run can then be expressed as $1 - (1 - \omega^l)^N$. Consequently, the expected number of iterations can be determined as follows:

\begin{equation}
\label{Iter}
\mathbb{E}[{N}] \approx \frac{\log_{10}(1-\textrm{Pr}(\textrm{sucsess}))}{\log_{10}(1-\omega^l)}.
\end{equation}

 Based on equation~(\ref{Iter}), when the success probability, $\textrm{Pr}(\textrm{sucsess})$, is set as a constant $pr$, the expected number of iterations increases as the subset sample size $l$ and the percentage of outliers increase. This increase in iterations adversely impacts the complexity. As a result, for large-scale problems, this approach is not optimal or practical due to the excessively high number of required iterations.


}
\end{remark}

\subsection{Identifying the mixing matrix}
\label{IMM}

Algorithm~\ref{Alg 1} is designed to identify $c$ orthogonal complement subspaces, denoted as $\mathcal{P} \in \mathbb{R}^{m \times (m-k) \times c}$. Each column of $\textbf{A}$ (i.e., $\textbf{a}_q$) lies in the intersection of $f={{n-1}\choose{k-1}}$ subspaces spanned by columns of $\textbf{A}$ that involve $\textbf{a}_q$. Consequently, the mixing vector is orthogonal to the $f$ orthogonal complement subspaces. To address the second stage of UBI, the approach involves finding and clustering the normal vectors to the $f$-combination of $c$ orthogonal complement subspaces. This can be achieved by estimating the eigenspaces through eigenvalue decomposition (EVD) of the covariance matrix for each $f$ orthogonal complement subspace. The algorithm proceeds by calculating the minimum eigenvalues and their corresponding eigenvectors, sorting them, and selecting the $n$ eigenvectors corresponding to the $n$ minimum eigenvalues. The detailed procedure is presented in Algorithm~\ref{Alg 2}.
 
\setalgorithmicfont{\small}

  \begin{algorithm}
  \captionsetup{font=small}

\caption{Identification of the mixing matrix based on EVD } 

\label{Alg 2}
\begin{algorithmic}

   \STATE \textbf{Input:}~$\mathcal{P} \in{\mathbb{R}^{m \times (m-k) \times c}}$ (orthogonal complement subspaces identified using Algorithm \ref{Alg 1}).
      \STATE \textbf{Output:} $\hat{\textbf{A}}$: estimated mixing matrix.
 \STATE \textbf{1:} Find all possible $f$-combinations of the set $\mathbb{C}=\{1,\dots,c\}$, i.e., $\mathbb{G}=$ $\mathbb{C} \choose f $, where $f \leftarrow {{n-1}\choose{k-1}}$. $\textbf{G}(:,j) \leftarrow j^{th}$ $f$-combination of $\mathbb{C}$ where $j=1$ to $g=$ $c \choose f $.
\FOR{$j=1$ to $g$} \STATE {

\textbf{2:} $\mathcal{P}_{sel}\leftarrow \mathcal{P}(:,:,\textbf{G}(:,j))$.}

\STATE \textbf{3:} $\textbf{P}_{sel}$ $\leftarrow$ $[\mathcal{P}_{sel}(:,:,1),...,\mathcal{P}_{sel}(:,:,f)] \in{\mathbb{R}^{m \times f.(m-k) }}$ . 
\STATE \textbf{4:} Build $\textbf{R} \leftarrow \textbf{P}_{sel}\textbf{P}_{sel}^\textrm{T}$.
\STATE \textbf{5:} Apply EVD on $\textbf{R}$ and obtain the minimum eigenvalue, $\Lambda(j) \leftarrow \lambda_1$ and its corresponding eigenvector $\textbf{E}(:,j) \leftarrow \textbf{e}_1$.
\ENDFOR
\STATE \textbf{6:} Choose $n$ eigenvectors that correspond to $n$ minimum values of $\Lambda$. These $n$ eigenvectors constitute the estimated mixing matrix, $\hat{\textbf{A}}$. 
\end{algorithmic}
\end{algorithm}

To overcome the exponential computational cost associated with calculating all possible $f$-combinations of the set $\mathbb{C}=\{1,\dots,c\}$, an alternative algorithm with lower computational complexity is required to identify the mixing matrix. One such solution is the subspace selective search ($\textrm{S}^3$) algorithm, proposed in our previous work~\cite{eqlimi2015multiple}. In $\textrm{S}^3$, the mixing vectors are identified through a selective search process, detecting as few as $m$ subspaces. While $\textrm{S}^3$ exhibits high accuracy and speed in noiseless scenarios, it struggles in the presence of noise and outliers due to the requirement of defining multiple thresholds (see Algorithm 3 in~\cite{eqlimi2015multiple}).
 

In this paper, we address the challenge of capturing noisy scenarios and removing outlier data points by employing a RANSAC-based subspace search for the identification of the mixing matrix. The proposed approach, presented in Algorithm \ref{Alg 3}, provides an integrated solution to the UBI problem by leveraging a unified approach for both stages of the problem. Notably, Algorithm \ref{Alg 3} utilizes equations (\ref{EqFit})-(\ref{EqDis}) as the RANSAC functions with $l=m-1$.


\setalgorithmicfont{\small}

\begin{algorithm}
\captionsetup{font=small}

\caption{Identification of the mixing matrix by RANSAC-based subspace search} 
\label{Alg 3}
\begin{algorithmic}

   \STATE \textbf{Initialization:} $Th_2$ (distance threshold for RANSAC), $l=m-1$ (the number of selected columns), $e \leftarrow 0$, $\hat{\textbf{a}} \leftarrow \emptyset$, $n^{\hat{A}} \leftarrow 0$ and $Th_3$ (distance threshold for the generative  clustering).
   \STATE \textbf{Input:}~$\mathcal{P} \in{\mathbb{R}^{m \times (m-k) \times c}}$ (orthogonal complement subspaces).
      \STATE \textbf{Output:} $\hat{\textbf{A}} \in{\mathbb{R}^{m \times n}}$ (estimated mixing matrix).
 \STATE \textbf{1:} Assign an initial set of inlier indices, $\mathbb{J}\leftarrow \{1, \dots, c\}$ ($\set{j \in \field{N} \given 1\leq j \leq c }$).
  \STATE {\textbf{2:} $\textbf{P}$ $\leftarrow$ $[\mathcal{P}(:,:,1),...,\mathcal{P}(:,:,c)] \in{\mathbb{R}^{m \times (m-k).c }}$.}  
\WHILE{$n^{\hat{A}}<n$  }
\STATE \textbf{3:} Apply RANSAC method to $\textbf{P}(:, \mathbb{J})$ following equations (\ref{EqFit})-(\ref{EqDis}) and obtain the indices set of the inlier data points $\leftarrow \mathbb{Y}$.
 \IF{$\mathbb{Y} \neq \emptyset$}
 \STATE {\textbf{4:} $e \leftarrow e+1$.}
\STATE {\textbf{5:} Decompose $\textbf{P}(:,\mathbb{Y})$ to $\textbf{U}\boldmath{\Sigma}\textbf{V}^\textrm{T}$ using economy-size SVD and find the normal vector,
$\textbf{P}_A(:,e) \leftarrow \textbf{U}(:,m)$ where $\textbf{U}(:,m)$ corresponds to minimum singular value.}
\FOR{$i=1$ to $e$} \STATE{ \textbf{6:} If $e=1$ then $\hat{\textbf{a}}_1 \leftarrow \textbf{p}^A_i=\textbf{P}_A(:,e)$ and $n^{\hat{A}} \leftarrow 1$ otherwise find absolute cosine distance values between $\textbf{p}^A_i=\textbf{P}_A(:,e)$ and $\forall$~$\hat{\textbf{a}}_j$ employing equation~(\ref{ACD}), where $j=1$ to $\hat{n}$ and $\hat{n}$ is the number identified mixing vectors.} 
\STATE{ \textbf{7:} Obtain the minimum distance,  $D=\textrm{min(ACD)}$, $\hat{\textbf{a}}_m$ $\leftarrow$ the closest vector to $\textbf{p}^A_i$, and sign of $\langle\hat{\textbf{a}}_m,\textbf{p}^A_i\rangle \leftarrow s$.} 
 
  \IF{$D<Th_3$} \STATE{\textbf{8:}  Update $\hat{\textbf{a}}_m$ by $\hat{\textbf{a}}_m \leftarrow \frac{\hat{\textbf{a}}_m+s.\textbf{p}^A_i}{2}$.}
  \ELSE \STATE{\textbf{9:}} Generate a new mixing vector, $\hat{\textbf{a}}_{\hat{n}+1} \leftarrow s.\textbf{p}^A_i$; $n^{\hat{A}} \leftarrow \hat{n}+1$.
\ENDIF

\ENDFOR
\ENDIF
\STATE \textbf{10:}  $\mathbb{J} \leftarrow$  a random permutation of the integers from $1$ to $c$ inclusive.

\ENDWHILE
\end{algorithmic}
\end{algorithm}

In Fig.~\ref{OverviewFig} (right block), the RANSAC method takes the orthogonal complement subspaces obtained from Algorithm \ref{Alg 1} as input. By utilizing the RANSAC-based selective search instead of exploring all possible $f$-combinations of $\mathbb{C}$, the computational cost of Algorithm~\ref{Alg 2} is effectively mitigated, leading to faster identification of the mixing vectors. In each iteration, the normal vectors of the inlier data are compared with those identified in previous iterations. If a new vector is discovered, a new cluster and vector are generated. We employ a generative clustering method proposed in~\cite{eqlimi2015multiple}, which utilizes the absolute cosine distance (ACD) to measure the distance between vectors.


\begin{equation}
\label{ACD}
\textrm{ACD}(\textbf{a}_j,\textbf{p}^A_i)= 1- \textrm{cos}(\theta),~\textrm{cos}(\theta)=\frac{\textbf{a}_j^\textrm{T}\textbf{p}^A_i}{||\textbf{a}_j||~||\textbf{p}^A_i||},
\end{equation}

\noindent
where $\textbf{a}_j$ represents the $j^{th}$ mixing vector and $\textbf{p}^A_i$ denotes the $i^{th}$ normal vector of the inlier subspace (step $6$ in Algorithm~\ref{Alg 3}). The three-stage process, consisting of RANSAC, normal vector finding, and generative clustering, is repeated until all $n$ mixing vectors are successfully identified.

\section{Simulations and Results}
\label{ResSec}

The method proposed in~\cite{naini2008estimating, amini2006fast} utilizes the Gaussian mixture model (GMM) to generate the sources, allowing for sparser vectors such as $1$-sparse in addition to $k$-sparse vectors. This flexibility simplifies the problem by leveraging sparsity. In contrast, our approach tackles a more challenging scenario where exactly $k=m-1$ sources are active at each time instant. However, the inactive components are not constrained to be zero, and instead, Gaussian noise with a small standard deviation $\sigma_{off} \ll 1$ is considered over the inactive elements.

In the first experiment, we consider a noiseless UBI problem with a mixing matrix described in~\cite{washizawa2006line}. The parameters are set as $m=3$, $n=5$, $k=2$, $T=2000$, and $\sigma_{off}=0$. We evaluate the identification error using the biased angle sum (BAS)~\cite{he2006k} and the Frobenius norm of the error matrix~\cite{washizawa2006line}. The BAS measures the sum of deviation angles ($\sbullet  \si{\degree}$) between the original and estimated mixing vectors, calculated as follows:
\begin{equation}
\textrm{BAS}(\textbf{A},\hat{\textbf{A}})=\frac{180}{\pi}\sum_{i=1}^{n}\arccos(\frac{\langle\textbf{a}_i,\hat{\textbf{a}}_i\rangle}{||\textbf{a}_i||~||\hat{\textbf{a}}_i||})
,
\end{equation}

\noindent where  $\textbf{A}=[\textbf{a}_1,\dots, \textbf{a}_n]$ and  $\hat{\textbf{A}}=[\hat{\textbf{a}}_1,\dots, \hat{\textbf{a}}_n]$ are the original and the estimated (optimally ordered~\cite{de2006second}) mixing matrices, respectively. 

Based on the results presented in Table~\ref{Exp1}, it is evident that the proposed algorithm surpasses both state-of-the-art algorithms in terms of average identification error and running time across $100$ trials.





\begin{table}[H]

\captionsetup{font=small}

\caption{Performance comparison of  the proposed algorithm and existing methods for the UBI problem presented in~\cite{washizawa2006line}. }\vspace{-4pt}
	\centering
	\scalebox{0.79}{%
	\begin{tabular}{ccclcccccc}\hline\hline
Method & $\textrm{BAS}$~[$\sbullet  \si{\degree}$]$ $ &  $||\textbf{A}-\hat{\textbf{A}}||_F$ & $\textrm{Time}~[\si{s}]$ 
\vspace{3pt}
\\ \hline
\vspace{3pt}
Adaptive $k$-plane clustering~\cite{washizawa2006line,washizawa2008sparse} & $    0.2088
$   & $0.2018$ &$3.31$\\
\vspace{3pt}
 Partial $k$-subspace clustering~\cite{naini2008estimating}  &  $ 0.0130$   & $0.0061$ &$7.33$ \\
Proposed Algorithm (~\ref{Alg 1}+~\ref{Alg 3})  & $1.2 \times 10^{-5}$  & $5.9\times 10^{-6}$ & $0.10$ \\
	\end{tabular}}
	\vspace{-8pt}	
	\label{Exp1}

\end{table}

\normalsize



The second experiment evaluates the performance of the algorithm for different values of $[m,n]$ and varying levels of $\sigma_{off}$ when $k=m-1$. The mixing vectors are randomly generated from a normal distribution and normalized to have a unit norm.

\begin{figure}[H]

    \centering
    \subfloat[]{{\includegraphics[height=4.2cm,width=4.7cm]{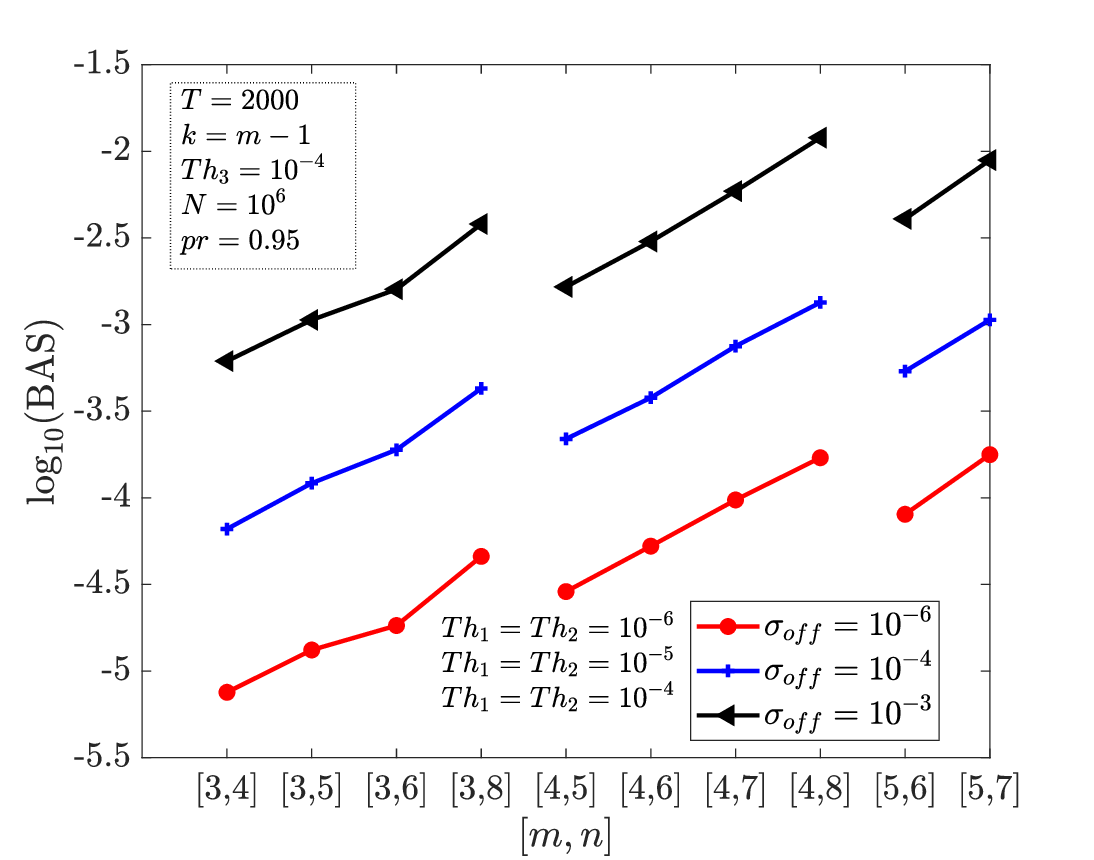} }
        \label{BAS_a}
}
    \subfloat[]{{\includegraphics[height=4.2cm,width=4.7cm]{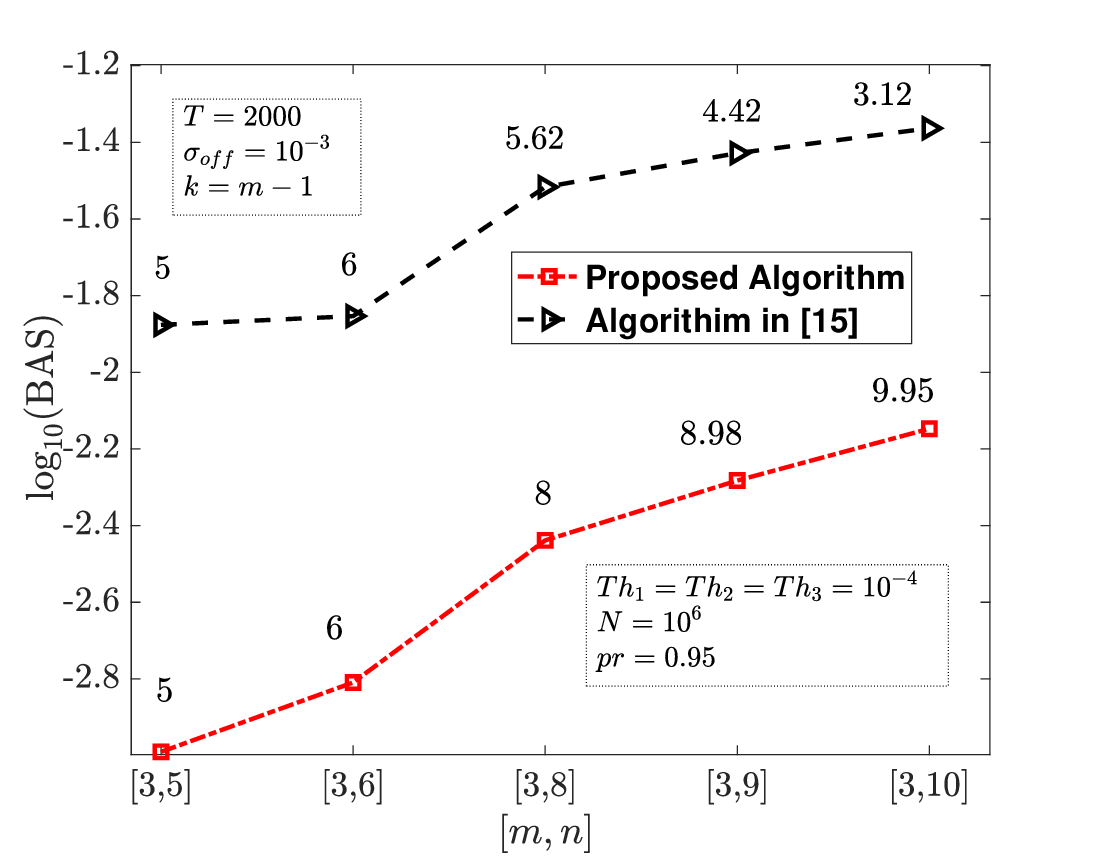} }

     \label{BAS_b}}
    \captionsetup{belowskip=0pt}

    \captionsetup{font=small}
    \caption{a) Identification error of our algorithm across different $[m,n]$ values for varying levels of $\sigma_{off}$; b) Comparison of our algorithm with the algorithm in~\cite{naini2008estimating} across different $[m,n]$ values for the noisy case with $\sigma_{off}=0.001$. The values above the marker indicate the average number of accurately identified vectors.
}
\end{figure}

Fig.~\ref{BAS_a} shows the average identification errors on a logarithmic scale for different $[m,n]$ values and three levels of $\sigma_{off}$ over 100 trials. The results demonstrate the robustness of our algorithm in the presence of noise on inactive sources. As $n$ and $\sigma_{off}$ increase, the identification error also increases, but the orientations of the identified vectors remain accurate (with $\textrm{BAS}<0.01 \si{\degree}$). In Fig.~\ref{BAS_b}, we compare our algorithm with the algorithm presented in~\cite{naini2008estimating} for the noisy case with $\sigma_{off}=0.001$. To ensure a fair comparison, we only consider cases where the initial parameters of the algorithm in~\cite{naini2008estimating} are properly selected. The results show that our algorithm achieves more accurate identification of the mixing vectors. It is important to note that the BAS calculation in Fig.~\ref{BAS_b} excludes inaccurately identified vectors (with $\textrm{deviation angle}\geq 0.1\si{\degree}$). The values above the markers represent $\hat{n}$, which is the average number of accurately identified vectors out of 100 trials. Unlike the algorithm in~\cite{naini2008estimating}, our proposed algorithm is capable of identifying almost all mixing vectors accurately. Based on these results, we can conclude that our algorithm offers higher identification accuracy, less sensitivity to initial parameters, and reduced running time.

 \vspace{-5mm}





\section{Discussion and Conclusions }
\label{ConSec}
In general, $k$-SCA methods, which consider $k \leq m-1$ (where $m$ is the number of sensors), are more constrained compared to $\ell_1$-minimization and overcomplete dictionary learning methodologies~\cite{donoho2003optimally,aharon2006uniqueness} under the assumption of $k < \frac{\textrm{spark}(\textbf{A})}{2}$. In this paper, we propose a new $k$-SCA algorithm for identifying the mixing matrix using the Gram-Schmidt and RANSAC approaches. Our algorithm outperforms existing methods for two main reasons. First, unlike single dominant component-based methods~\cite{bofill2001underdetermined,abrard2005time,reju2009algorithm,thiagarajan2013mixing} which fail when there are insufficient $1$-sparse sources and highly sparse components, our algorithm performs well when $k=m-1$ sources are active at each time instant. This means that our algorithm can handle scenarios where most sources are active simultaneously. Second, unlike $k$-hyperplane~\cite{washizawa2006line} and $k$-EVD~\cite{he2006k} clustering methods that rely on the normal vector, our algorithm estimates orthogonal complement subspaces, enabling the handling of multiple dominant SCA. Additionally, it exhibits relative robustness to noise in inactive sources due to the RANSAC process.

In the second step of our algorithm, we propose a method similar to subspace identification to avoid combinatorial explosion. However, this method is not suitable for large-scale problems within a desirable time frame due to the exponential growth in RANSAC iterations ($N$). Consequently, solving large-scale $k$-SCA problems remains an open challenge. Additionally, our method relies on a few parameters, such as $Th_1$ and $Th_2$, which need adaptive estimation based on $\sigma_{off}$ and the number of subspaces. One potential solution to overcome these limitations is to utilize an optimization method for estimating these thresholds. As a potential application, our k-SCA algorithm may extract k-dimensional subspaces in electroencephalography (EEG) microstate analysis, enabling the capture of complex interactions among brain regions and enhancing our understanding of brain dynamics and functional connectivity~\cite{eqlimi2020eeg,eqlimi2022exploring}.

\bibliographystyle{IEEEtran}         
\bibliography{Sparse-RANSAC-UBI-SPL-Main.bib}
\end{document}